\documentclass[onecolumn,11pt]{article}
\usepackage[top=1in, bottom=1in, left=1in, right=1in]{geometry}
\usepackage{bm,amsmath,amsfonts,amscd,amssymb}
\usepackage{graphicx}
\usepackage{url}
\usepackage{caption}
\usepackage{color}
\usepackage{mathtools}
\usepackage{setspace}
\usepackage[numbers,sort&compress]{natbib}
\usepackage{framed}
\usepackage{enumitem}
\usepackage{newtxtext}
\usepackage{newtxmath}
\usepackage{hyperref}
\hypersetup{
  colorlinks = true,
  urlcolor  = blue,
  citecolor = black,
}

\newcommand{\bqf}{\boldsymbol{q}^f}
\newcommand{\bqs}{\boldsymbol{q}^s}

\newcommand{\bx}{\boldsymbol{x}}

\newcommand{\by}{\boldsymbol{y}}

\newcommand{\bC}{\boldsymbol{C}}

\newcommand{\bI}{\boldsymbol{I}}

\newcommand{\RomanNumeralCaps}[1]
\linenumbers

\newcommand\blfootnote[1]{%
  \begingroup
  \renewcommand\thefootnote{}\footnote{#1}%
  \addtocounter{footnote}{-1}%
  \endgroup
}

\title{\LARGE{\vspace{-.55in}\textbf{Network-theoretic modeling of fluid-structure interactions}}\vspace{-.175in}}
\title{\vspace{-.55in}{\fontsize{16}{16}\selectfont \textbf{Network-theoretic modeling of fluid-structure interactions}}\vspace{-.15in}}

\author{\normalsize{Aditya G.~Nair$^{*}$, Samuel B.~Douglass \& Nitish Arya}\\
\footnotesize{Department of Mechanical Engineering, University of Nevada, Reno, NV 89557}\\
}
\date{}

\begin{document}
\maketitle

\blfootnote{$^*$ Corresponding author (adityan@unr.edu).}
\vspace{-.2in}

\begin{abstract}
The coupling interactions between deformable structures and unsteady fluid flows occur across a wide range of spatial and temporal scales in many engineering applications.
These fluid-structure interactions (FSI) pose significant challenges in accurately predicting flow physics.
In the present work, two multi-layer network approaches are proposed that characterize the interactions between the fluid and structural layers for an incompressible laminar flow over a two-dimensional compliant flat plate at a 35-degrees angle of attack. 
In the first approach, the network nodes are formed by wake vortices and bound vortexlets, and the edges of the network are defined by the induced velocity between these elements.
In the second approach, coherent structures (fluid modes), contributing to the kinetic energy of the flow and structural modes, contributing to the kinetic energy of the compliant structure constitute the network nodes.
The energy transfers between the modes are extracted using a perturbation approach. 
Furthermore, the network structure of the FSI system is simplified using the community detection algorithm in the vortical approach and by selecting dominant modes in the modal approach. 
Network measures are used to reveal the temporal behavior of the individual nodes within the simplified FSI system.
Predictive models are then built using both data-driven and physics-based methods.
Overall, this work sets the foundation for network-theoretic reduced-order modeling of fluid-structure interactions, generalizable to other multi-physics systems.
\end{abstract}

\section{Introduction}

Fluid-structure interactions (FSI) occur in many engineering applications and over many spatial and temporal scales from aircraft and buildings to heart valves and insect wings. 
In fact, any compliant structure immersed in a fluid flow result in fluid-structure interaction.
These interactions are often transitory in nature and lead to the rich dynamical behavior of the fluid and structural components. 
For flight systems with compliant wings, the structure can extract energy from the air stream leading to an unstable self-excited vibration called flutter, which is not only difficult to predict but can have catastrophic effects such as potential structural failure \citep{wright2008introduction, mittal2004flutter, shoele2016flutter}. 
Due to their design characteristics, slender and High Altitude Long Endurance (HALE) aircraft wings are more prone to experiencing this self-excited vibration, posing significant challenges in their safe and efficient operation.
\citep{patil2001nonlinear,d2016high,fladeland2019demonstrating}.
The situation is similar for wind turbines, where increasing the aspect ratios driven by increases in turbine name-plate capacity leads to a higher likelihood of flutter \citep{enevoldsene2019xamining}.
Furthermore, as the utility of a wind turbine is to extract energy from the wind, any energy lost to or because of blade distortion is energy that could have been used to turn the generator, hindering the efficient extraction of power from the wind.
Active flutter alleviation systems which take advantage of the knowledge of the system interactions are of significant interest as they provide the potential for significant weight savings when compared to traditional flutter-resistant structures \citep{livne2018aircraft}. 

Interest in FSI extends to smaller scales as well.
Agile natural flyers such as insects and birds are able to maneuver in unsteady aerodynamic environments.
Because many insects are unable to fully articulate their wings, wing compliance plays a crucial role in the generation of flight forces \citep{dickinson1992directional, young2009details, mountcastle2010aerodynamic, mountcastle2010vortexlet}.
This provides insights into the design and control of autonomous flight vehicles \citep{colmenares2015compliant, li2018review}, a topic of tremendous engineering interest.

Because of the prevalence of FSI and the potential for catastrophic phenomena, significant effort has been made in modeling and predicting their behavior \citep{dowell2001modeling}.
Efforts have ranged from simple analytical methods and semi-empirical equations of prediction \citep{hodges2011introduction} to computationally-intensive high-fidelity numerical simulations \citep{brown2012accuracy, shinde2019transitional}. 
Perhaps the most commonly used analytic approach is Theodorsen's model which was motivated by the importance of understanding wing vibrations and flutter in the early years of flight.
Improvements have been made to the model in recent years including semi-empirical formulations \citep{mahajan1993semi}, state-space models \citep{brunton2013empirical, hickner2022data} and insights from careful experiments \citep{hessenthaler2017experiment,gomes2011experimental}.
Vortex methods can be coupled to low-fidelity structural models to build fast solvers, but their speed comes at the expense of ignoring viscous and compressible effects in the flow \citep{eberle2014fluid, ramesh2014discrete, kuzmina2018numerical, fasel2022flexwing}.
In high-fidelity simulations, it is common to employ partitioned solvers for each component physics which are then coupled using implicit or explicit coupling schemes \cite{zheng2010coupled}. 
This often increases the computational cost and the likelihood of numerical stability issues compared to simulating each system separately \citep{bungartz2016precice, landajuela2017coupling, formaggia2007stability}.
In the present work, we propose two separate mathematical frameworks for modeling coupled fluid-structure systems with a specific focus on capturing the interactions between the two systems.

Network science and graph theory provide a concise and powerful mathematical framework for the interactions between actors within a system.
In a network representation of a system, actors within the system are represented as nodes, and the interaction between the nodes (actors) is represented by edges connecting them.
Mathematically, a network is represented by a graph $\mathcal{G} = \left ( \mathcal{V}, \mathcal{E}, \mathcal{W} \right )$ where nodes $\mathcal{V}$ are connected via edges $\mathcal{E}$, each with an associated edge weight $\mathcal{W}$ \citep{newman2010networks}.

Network science has found extensive usage in social sciences \citep{borgatti2009network, ferrara2012large, campedelli2019complex}, biology \citep{barabasi2004network, gosak2018network}, computer science \citep{deo2017graph}. In the realm of physics and engineering, it has been employed to explore the interaction between global climate modes \cite{harries2021dynamic}, study of turbulent and vortex dominated flows \citep{iacobello2021review, taira2022network} and the study of thermoacoustic combustion instabilities \citep{sujith2020complex}.
This work aims to build a scaffolding of network-based approaches for modeling FSI systems. 
The advantage of the network approach is that it naturally allows for the incorporation of physics-based insights in data-driven system identification strategies \citep{kou2021data} such as those based on proper orthogonal decomposition \citep{berkooz1993proper}, dynamic mode decomposition \citep{schmid2010dynamic}, and eigensystem realization algorithm \citep{juang1985eigensystem}.
The approach also naturally lends itself to the systematic reduction of the physical system via community detection \citep{fortunato2010community, meena2018network, meena2021identifying} and graph sparsification algorithms \citep{spielman2008graph, nair2015network}, along with identifying the key nodes controlling the system dynamics \citep{taira2016network, yeh2021network}.

In this work, we present two approaches for modeling FSI using a network-based framework. 
The first approach characterizes the vortical interactions in FSI with the network nodes in the fluid and structure domains defined by discrete point vortices. 
The edge weights are based on the induced velocity of these point vortices \citep{nair2015network}.
We also introduce a modal network representation of FSI where the network nodes are given by coherent spatial modes of the unsteady fluid flow and velocity modes of the structure.
Data collected from perturbations of the structural modes are used to determine the interaction strengths (edge weights) between the nodes \citep{nair2018networked}.
Both approaches not only highlight interactions within each component part of FSI but also extracts the cross-coupling interactions in the form of a multilayer network \citep{kivela2014multilayer}, i.e., one network layer for the fluid and one for the structure.

We demonstrate the network modeling approaches for a two-dimensional laminar flow over a compliant flat plate at an angle of attack $\alpha = 35^\circ$. 
A similar problem was investigated in the work by Hickner et al.\cite{hickner2022data} for developing data-driven system identification models. 
However, system identification in that work was restricted to flows in the steady regime with the angle of attack below the critical angle of attack of $\alpha = 27^\circ$. 
In this work, we analyze the FSI interactions in the unsteady regime as well as those on the introduction of large disturbances to the flow caused by gust encounters. 
We discuss the numerical setup and methods in section \ref{methods}, results in section \ref{results}, and offer concluding remarks in section \ref{conclusion}.

\section{Methods}
\label{methods}

\subsection{Direct numerical simulation}
\label{setup}

We perform direct numerical simulations of two-dimensional incompressible laminar flow over a thin deforming flat plate of length $c$ at an angle of attack of $\alpha = 35^\circ$. 
These simulations are performed using the strongly-coupled immersed boundary method \citep{goza2017strongly, taira2007immersed}. 
The solver uses a multi-domain technique to accelerate the computations \citep{colonius2008fast}.
We use five grid levels with the innermost domain fixed at $-0.2 \le x/c \le 1.8$ and $-1 \le y/c \le 1$, with a grid spacing of $\triangle x/c \approx 0.0077$. 
We carry out a detail grid convergence study for the present study in Appendix A.
Uniform flow with free-stream velocity $U_\infty$ is prescribed at the far-field boundaries.
An explicit Adam-Bashforth method is used for the discretization of the advection term and an implicit Crank-Nicolson scheme is used for the viscous terms of the governing equations.
The flat plate is evolved using the Euler--Bernoulli equation with a co-rotational finite element discretization. 
Such a co-rotational form allows for large displacements of the structure. 
The plate is discretized into $65$ elements ($66$ points) with the leading edge pinned at $(x/c,y/c) = (0,0)$.

The FSI system is characterized by three non-dimensional parameters: Reynolds number $Re = U_\infty c/\nu$, mass ratio $M_{\rho} = \frac{\rho_s h}{\rho_f c} = 3$, and bending stiffness $K_{B} = \frac{EI}{\rho_f U_{\infty}^2 c^3}$.
Here, $\nu$ is the kinematic viscosity, $\rho_s$, and $\rho_f$ are the densities of the structure and fluid, respectively. Also, $h$ is the thickness, $E$ is Young's modulus, and $I$ is the second area moment of inertia of the plate. 
We fix $Re = 100$ and $M_\rho = 3$, unless otherwise stated. 
Data from numerical simulation of three different bending stiffness $K_B = \{0.15625, \ 0.3125, \ 0.625\}$ are collected \citep{combes2003flexural, hickner2022data}. 

\begin{figure}
    \includegraphics[width=1.00\textwidth]{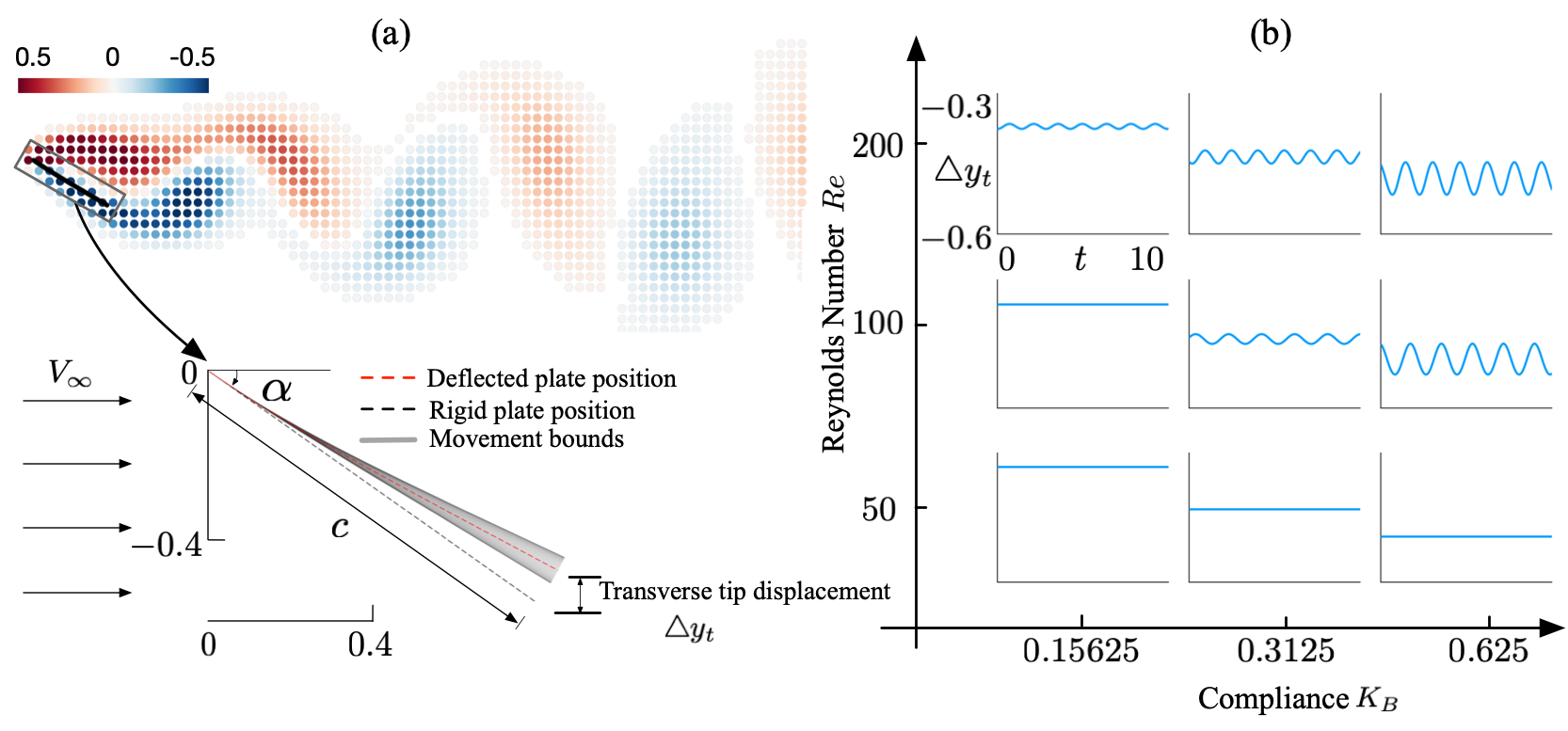}
    \caption{Direct numerical simulation of 2D laminar flow over a compliant flat plate of length $c$ at an angle of attack $\alpha = 35^\circ$: (a) Vorticity snapshot (top) and the numerical setup of the flat plate (bottom). (b) The transverse tip displacement across different Reynolds number $Re$ and bending stiffness $K_B$.}
    \label{fig:setup}
\end{figure}

We show a snapshot of vorticity in the top panel of Figure \ref{fig:setup}(a) and the flow field parameters and domain setup of the structure in the bottom panel. 
The setup also highlights the position of a rigid body at an angle of attack of $\alpha = 35^\circ$ along with the deflected position for a complaint case. 
The transverse tip displacement $\triangle y_t$ is always negative for the cases considered in this work. 
By convention, we consider positive transverse tip displacement of the trailing edge when the plate pitches down compared to the rigid plate position. 
We also show the tip displacements for three different Reynolds numbers and the three bending stiffnesses in Figure \ref{fig:setup}(b).
With the choice of the parameters considered, the fluctuation of the tip displacement increases with $Re$ and $K_B$ and the mean tip displacement increases with $K_B$. 

\subsection{Fluid-structure vortical interaction networks} 
\label{fsvortnet}
Due to the different physical nature of the fluid and structural components and their governing dynamics, we model each of them into separate vortical network layers and then combine them later to form the multilayer network.  
To construct the network, we collect snapshots of data from direct numerical simulations of the FSI problem as described in section \ref{setup}.
We then convert this data to a network-based representation as illustrated below. 

The fluid network layer is created with the method already described in previous work by Taira et al. \cite{taira2016network}, and Meena et al. \cite{meena2018network}.
Here, spatial grid points serve as network nodes.
The strength of each node is determined by the circulation $\gamma_{i}^f$ corresponding to the grid cell it represents.
The superscript $f$ indicates the nodes in the fluid layer. 
We only consider nodes in the fluid layer with vorticity values greater than 1\% of the maximum vorticity of the flow.
The influence of these nodes on each other is given by their induced velocity. 
The velocity induced by node $j$ on node $i$ is given by $u^f_{i\leftarrow j}$ and helps define the fluid layer network $\mathcal{G}^f$.
The node $i$ does not induce velocity on itself. 
The network can be neatly summarized with an adjacency matrix $\mathbf{A}^f$ as 
\begin{equation}
    A_{ij}^f = \begin{cases}
        u^f_{{i}\leftarrow {j}} & \text{if } i \neq j \in~\text{fluid layer} \\
        0 &\text{otherwise} 
        \end{cases}
        \quad\quad \text{where} \quad\quad
        u_{i\leftarrow j}^f = \frac{\gamma_{j}^f}{2\pi \vert \mathbf{r}^f_j - \mathbf{r}^f_i \vert}.
\label{eqn:vortical_netwrok_definition}
\end{equation}
where $\mathbf{r}^f$ is the location of the grid cell. 
The above definition leads to a weighted, directed network. 
Here, we consider $N$ fluid nodes after vorticity thresholding to construct the adjacency matrix. 

Vorticity is not a natural quantity to consider when dealing with structural mechanics.
However, we can represent the structure as a vortex line element formed of bound vortexlets, following the method by Mountcastle et al. \cite{mountcastle2010vortexlet}.
In this formulation, for a flat plate, the flow separates tangentially from the trailing edge, enforcing the Kutta condition, no-penetration boundary condition, and Kelvin’s circulation theorem.
We define $n$ control points on the structure co-located with every bound vortexlet. 
To calculate the strength of bound vortexlets $\gamma_i^s$ (superscript $s$ indicates the nodes in the structural layer) corresponding to each point on the structure, a linear system of equations is solved using the position and velocity of the structure and strength (circulation) of the fluid nodes above obtained from high-fidelity numerical simulations as
\begin{equation}
   \begin{bmatrix}
   \gamma_{1}^s\\
   \gamma_{2}^s\\
   \vdots\\
   \gamma_{n}^s
   \end{bmatrix} =    \begin{bmatrix}
   M_{s1,p1} & \dots & M_{sn,p1} \\
   M_{s1,p2} & \dots & M_{sn,p2} \\
   \vdots& \ddots & \vdots\\
   1 & \dots & 1
   \end{bmatrix}^{-1}  \left ( 
   \begin{bmatrix}
   \vec v^p_{1} \cdot \hat n^p_{1}\\
   \vec v^p_{2} \cdot \hat n^p_{2}\\
   \vdots\\
   0
   \end{bmatrix}\right .\\ 
    -   \left.\begin{bmatrix}
   M_{f1,p1} & \dots & M_{fN,p1}  \\
   M_{f1,p2} & \dots & M_{fN,p2} \\
   \vdots& \ddots & \vdots\\
   1 & \dots & 1
   \end{bmatrix} 
   \begin{bmatrix}
   \gamma_{1}^f\\
   \gamma_{2}^f\\
   \vdots\\
   \gamma_{N}^f
   \end{bmatrix}\right)
\label{eqn:bound_vortexlet}
\end{equation}
where the mass matrix is defined as 
\vspace{0.2cm}
\begin{equation}
    M_{s(f)i,pj} = \left [ \frac{-(y^p_{j} - y^{s(f)}_i)}{2\pi (r^2 + \delta^2)}, \frac{(x^p_{j} - x^{s(f)}_i)}{2\pi (r^2 + \delta^2)} \right ].
    \label{eqn:mass_matrix}
\end{equation}
Here, ($x^p_i, y_{i}^p$), $v^p_{i}$, $\hat n^p_{i}$,  are the position, velocity, and normal vector of each control point along the body, respectively. 
Also, ($x^{s(f)}_{j}, y^{s(f)}_{j}$) is the position of bound (fluid) vortexlet $j$,  $r^2 = (x - x_i)^2 + (y - y_i)^2$ and $\delta$ is a smoothing parameter preventing a divide by zero when $r = 0$.  
We chose $\delta = 0.001$ as the smoothing parameter. 
Experimentation shows that this value was sufficiently small so as to not significantly impact the results and obtain consistent vortical strengths compared to other values.
The nodes of the structural layer consist of bound vortexlets. 
Once again, we use induced velocity to quantify the interactions between the bound vortexlets  which leads to the adjacency matrix $\mathbf{A}^s$ given by
\begin{equation}
A_{ij}^s = \begin{cases}
        u^s_{{i}\leftarrow {j}} & \text{if } i \neq j \in~\text{structure layer} \\
        0 &\text{otherwise} 
        \end{cases}
        \quad\quad \text{where} \quad\quad
 u^s_{i\leftarrow j} = \frac{\gamma_j^s}{2\pi \vert\mathbf{r}^s_{j} - \mathbf{r}^s_{i}\vert},
 \end{equation}
where $\mathbf{r}^s$ are the location of points on the structure. 

An important measure that describes the global influence of the nodes in the network is the node degree or strength. 
The in-degree is defined as $s^{\text{in}}_i = \sum_{j = 1}^N A_{ij}^{s(f)}$, while the out-degree is given by $s^{\text{out}}_i = \sum_{i = 1}^N A_{ij}^{s(f)}$. 
The nodes with the maximum out-degree influence the network the most, while those with the maximum in-degree get influenced the most. 
With the fluid and structural layers defined, we reduce each network using community detection before combining them into a multilayer representation.

Community detection groups the nodes within a network to form distinct communities.
Nodes with a community have a higher density of interactions amongst themselves than with nodes in the other communities.
We utilize the Louvain algorithm \citep{traag2019louvain} to find communities that maximize modularity of the network \citep{newman2006modularity} defined as 
\begin{equation}
    Q = \frac{1}{2m}\sum_{ij}\left (A_{ij}^{s(f)} - \frac{s_i^{\text{in}}s_j^{\text{out}}}{2m}\right )\delta(C_i,C_j) \in [0, 1]
    \label{eqn:modularity}
\end{equation}
where $m$ is the number of nodes and $\delta$ is the Kronecker delta operating on the community labels $C_i$.
Modularity provides a measure of the relative connectedness of a group of nodes compared to their expected connectedness produced by a null model.
As the Louvain algorithm can only be applied to unsigned edge weights, we separate the fluid and structural network layers into ones that contain positive or negative edge weights and apply community detection.

\begin{figure}[h]
    \includegraphics[width=1.00\textwidth]{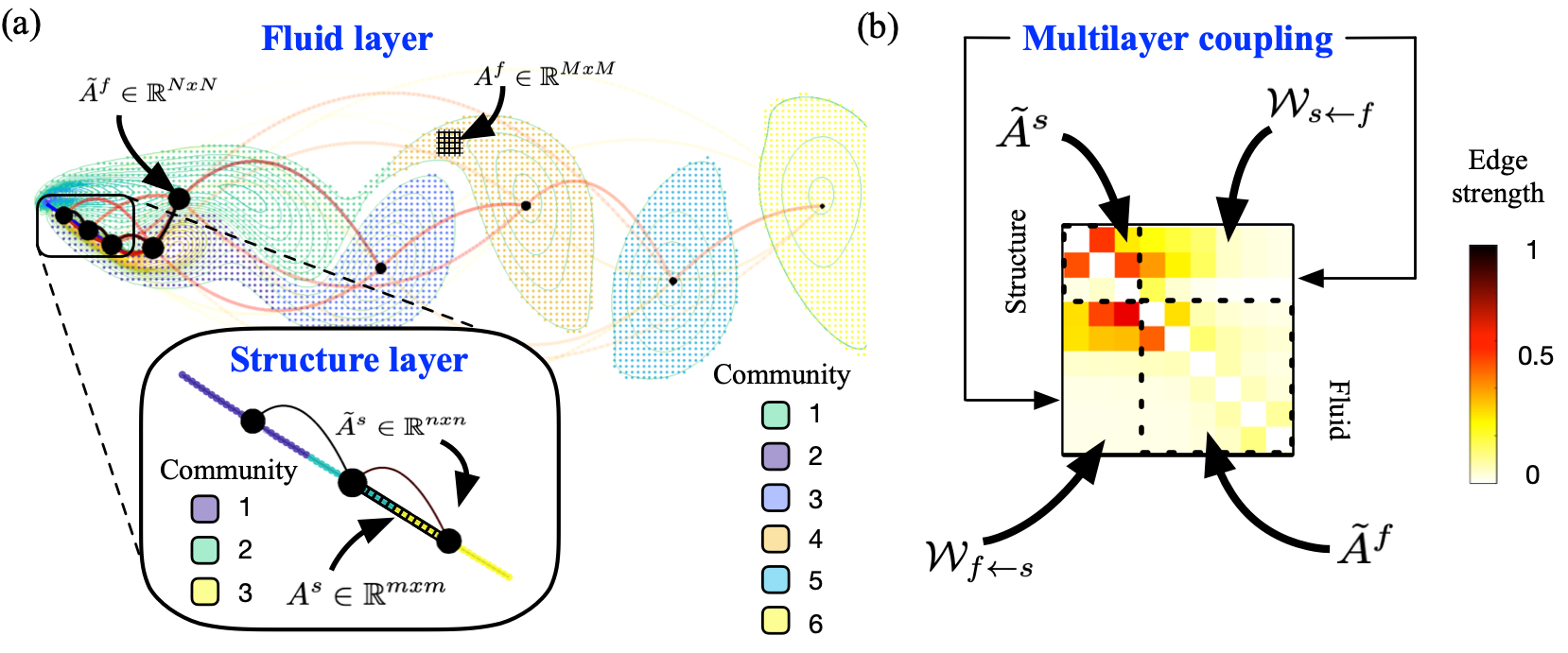}
    \caption{Fluid-structure vortical interaction network for 2D laminar flow over a flat plate ($M_\rho = 3$, $K_B = 0.3125$, $Re = 100$): (a) Community reduction of the fluid network layer and the structure network layer. (b) Supra-adjacency matrix containing edge weights for the structure layer (top main-diagonal block) and fluid (bottom main-diagonal block) and the inter-layer fluid-structure coupling and structure-to-fluid coupling on the off-diagonal blocks. The edge weights are normalized with respect to the maximum edge weight for visualization.}
    \label{fig:sparsification}
\end{figure}

The results of community detection applied to one snapshot of the flow field are shown in Figure \ref{fig:sparsification}(a). 
The community detection of the structural layer yields $n_c = 3$ communities while that of the fluid layer yields $N_c = 6$ communities.
For each community, we compute the community centroid shown by the filled black circles. 
The size of the circle indicates the node degree or strength of the community centroid. 
Through community reduction, we achieved a drastic reduction in the dimensionality of the FSI system from $n = 66$ to $n_c = 3$ for the structural layer and from $N = 67600$ to $N_c = 6$ for the fluid layer.

Using the community centroids identified above, we now are ready to define a community-reduced adjacency matrix for each layer as well as a combined multilayer adjacency matrix. 
Each community centroid $c_i$ has an associated strength $\gamma_{c_i}^{s(f)}$ and position $(x_{c_i}^{s(f)},y_{c_i}^{s(f)})$. 
The community-reduced adjacency matrix for the structural layer $\mathbf{\tilde{A}}^s$ and the fluid layer $\mathbf{\tilde{A}}^f$ are given by 
\begin{equation}
\begin{split}
\tilde{A}_{c_i, c_j}^s = \begin{cases}
        u^s_{{c_i}\leftarrow {c_j}} & \text{if } c_i \neq c_j \in~\text{structure layer} \\
        0 &\text{otherwise} 
        \end{cases}\\
\tilde{A}_{c_i,c_j}^f = \begin{cases}
        u^f_{{c_i}\leftarrow {c_j}} & \text{if } c_i \neq c_j \in~\text{fluid layer} \\
        0 &\text{otherwise}.
        \end{cases}
\end{split}
 \end{equation}
The combined network can be represented with a supra-adjacency matrix, $\mathcal{A}_{\alpha}$ that contains the adjacency matrices of both the fluid and structural layers along the block diagonal along with the inter-layer edge weight, $\mathcal{W}_{ij}$ at the off-block diagonal as
\begin{equation}
    \begin{aligned}
       \mathcal{A}_\alpha = 
       & \begin{bmatrix}
       \mathbf{\tilde{A}}^s & \mathcal{W}_{s\leftarrow f} \\
       \mathcal{W}_{f\leftarrow s} & \mathbf{\tilde{A}}^f 
        \end{bmatrix},
    \end{aligned}
    \label{eqn:supra_steps}
\end{equation}
where the inter-layer weights $\mathcal{W}_{s\leftarrow f}$ are the velocity induced by the fluid community centroids on the structural community centroids and $\mathcal{W}_{f\leftarrow s}$ are the velocity induced by the structural community centroids on the fluid community centroids. The supra-adjacency matrix is highlighted in Figure \ref{fig:sparsification}(b). The edge weights are normalized with the maximum edge weight for visualization. We see a lot of interactions among the structural nodes and the near wake fluid communities.

\subsection{Fluid-structure modal interaction network} 
\label{fsmodalnet}

To construct the modal interaction network, we perform proper orthogonal decomposition (POD) of the flow velocity field data obtained from the direct numerical simulations in section \ref{setup} to extract the most energetic coherent structures (modes).    
In this work, we only extract the modal network for the most compliant case of $K_B = 0.625$.
We employ the method of snapshots \citep{sirovich1987turbulence} to decompose the velocity fields $\bqf$ as
\begin{equation}
    \bqf(x, y, t) = \overline{\bqf}(x, y) + \sum_{j = 1}^{N} a_j^f(t) \boldsymbol{\phi}^f_j(x, y).
    \label{eqn:pod}
\end{equation}
where $N$ is the number of fluid modes, $\overline{\bqf}(x, y)$ is the mean flow, and $\boldsymbol{\phi}^f_j(x, y)$ are the fluid modes with temporal coefficients given by 
\begin{equation}
a_j^f(t) = \left <\bqf(x,y, t) - \overline{\bqf}(x,y), \boldsymbol{\phi}_j(x,y) \right >.
\end{equation} 
Here, $\left <\cdot, \cdot\right >$ stands for inner project.  We fix $N = 8$ to capture $99.9\%$ of the total energy of the fluid flow given by $\text{KE} = \bqf \cdot \bqf \approx \sum_{j=1}^{N} a_j^2/2 $.

Similarly, principal component analysis (PCA) is performed on the time series of x- and y-velocities, $\bqs  = (\dot \bx^s, \dot \by^s)$ of each of the structural elements to yield $p$ modes $\boldsymbol{\phi}^s$ and associated temporal coefficients $a_j^s$. We fix $p = 3$ to capture $99.9\%$ of the energetics of the structural deformations. 

Fluid flow modes appear in complex conjugate mode pairs. 
We combine these mode pairs to form an oscillator representation of their temporal dynamics as 
\begin{equation}
    z^f_m (t) = a^f_{2j-1} + ia^f_{2j} \approx r_m^f \exp(i\theta^f_m)
    \label{eqn:mode_pair_oscillator}
\end{equation}
where $j = 1,2,\dots,N/2$, $r_m = \| z_m \|$, and $\theta_m = \angle z_m$.
The oscillator number $m$ is denoted with Roman numerals to distinguish them from mode numbering $j \in {1,2,\dots,N}$.
We consider $M= N/2$ fluid oscillators.
The oscillator representation is akin to the polar decomposition of the temporal coefficients of the mode pairs. 
This helps in building a concise networked oscillator model, similar to the work of Nair et al. \cite{nair2018networked}. 
PCA of the structural velocity data does not yield modes in pairs as in the case of fluid data. 
To convert the temporal coefficients of the structures to oscillator representation, we perform the Hilbert transform \citep{feldman2011hilbert} of the temporal coefficients time-series data.
This transformation converts the real data sequence to an analytic signal (i.e. complex helical sequence), where the real part is the original data and the imaginary part is a version of the real sequence with a $90^\circ$ phase shift.  
The transformed series, which leads to structural oscillator representations, contain the same amplitude, frequency, and instantaneous phase information as the original signal.
The structural oscillators are given by $ z^s_m = r_m^s \exp(i\theta^s_m)$ corresponding to each temporal coefficient with $m = \text{I}, \text{II},\dots,p$. 

Once the fluid and structure oscillator representations are formed, we follow the procedure demonstrated in Nair et al. \cite{nair2018networked} to extract modal interaction networks. 
In Nair et al. \cite{nair2018networked}, impulse perturbations were introduced to the temporal coefficients of the fluid to induce interactions among the modes. 
However, this approach relies on exciting modes of the entire fluid domain, which is infeasible. 
In this work, impulse perturbations are introduced to the structural dynamics, which are both physically meaningful and realistic.
In particular, we add phase and amplitude impulse perturbations to the structural oscillators
The phase perturbations in the modes range from $-\pi$ to $\pi$ shifts in the phase of the modes relative to the baseline and the amplitude perturbation ranges from $0.1$ to $100\%$ of total kinetic energy.

To track the perturbations introduced and the spread among the fluid and structural modes, we normalize the oscillator representations for the fluid and structure modes to yield oscillator perturbations as $\xi^f_m = z_m^{f}/z_m^{f,b}$ and $\xi^s_m = z_m^{s}/z_m^{s,b}$, respectively. 
Here, $z_m^{f,b}$ and $z_m^{s,b}$ are the baseline fluid and structure oscillator trajectories, respectively.
Such a normalization yields zero perturbation amplitude at steady state and a finite steady-state phase shift. 
We collect data corresponding to three periods of baseline oscillation after the introduction of impulse perturbation.

Once the data for the perturbations are tracked and collected, we can form a multilayer network with structural and fluid oscillators as nodes.
Unlike the vortical network, the modal network lends itself to a combined representation automatically. 
A simple regression is performed on the perturbation datasets $\xi_m = \{\xi^s_m; \xi^f_m\}$ with $M+p$ oscillators. 
This results in a complex adjacency matrix for both the intra- and inter-layer interaction strengths between the structure and fluid oscillator layers as
\begin{equation}
\frac{d}{dt}
   \xi_m = \sum_{n = I}^{M+p} A_{mn} (\xi_n - \xi_m) = -\sum_{n = I}^{M+p} L_{mn}\xi_n
   \label{nom}
\end{equation}
where the complex adjacency matrix $\mathbf{A}$ and Laplacian matrix $\mathbf{L}$ are given by
\begin{equation}
    A_{mn} = \vert \omega_{mn} \vert\exp(i\angle \omega_{mn}),~~~~  L_{mn} = s_m^{\text{in}} - A_{mn}
    \label{eqn:oscillator_network}
\end{equation}
where $s_m^{\text{in}}$ is the standard in-degree. As the adjacency matrix is complex-valued, the magnitude of each edge $\vert\omega_{mn}\vert$ highlights the overall influence and the $\angle\omega_{mn}$ provides the phase relationship between the oscillators. To incorporate the insights from different impulse perturbation tests, we separate the data into training and test sets and perform model selection on the adjacency matrices obtained.   

\begin{figure}
    \includegraphics[width=1.00\textwidth]{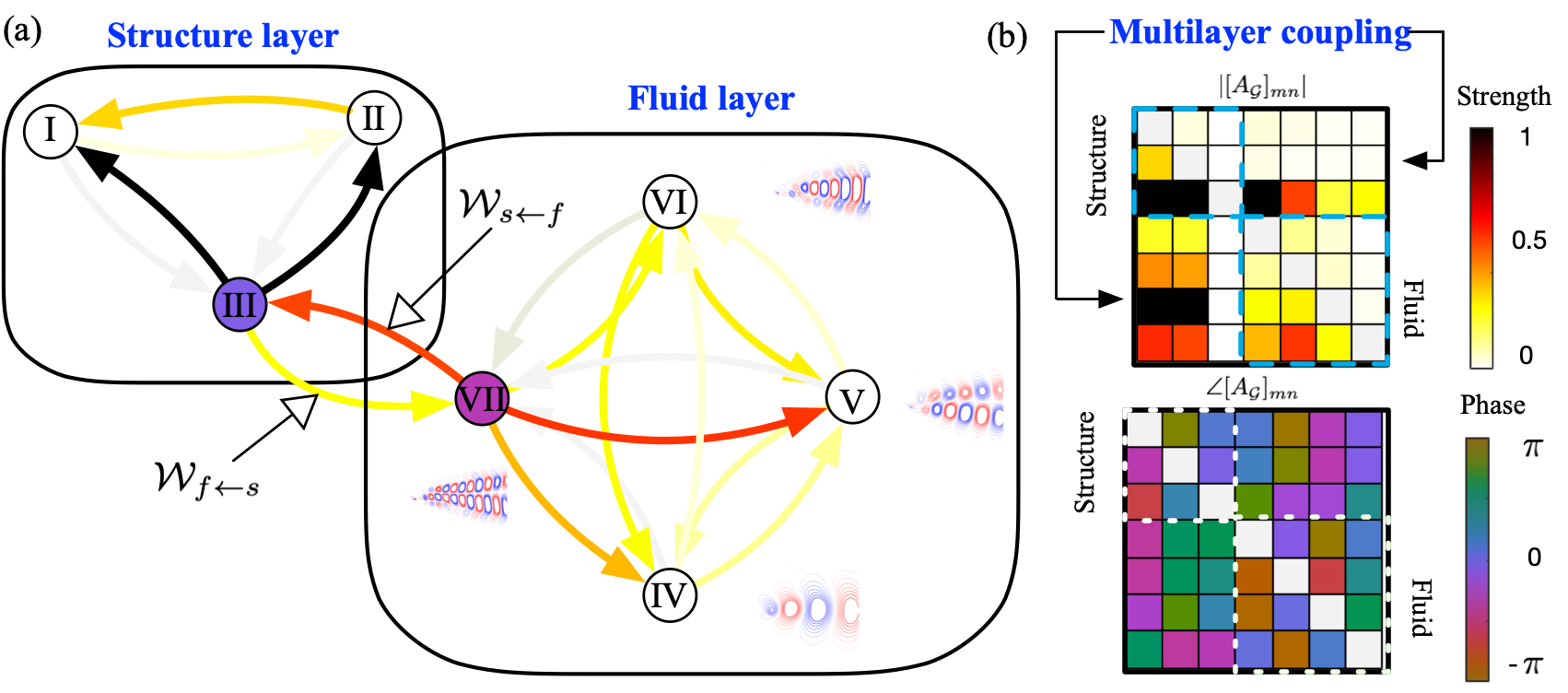}
    \caption{Fluid-structure modal interaction network for 2D laminar flow over a flat plate ($M_\rho = 3$, $K_B = 0.625$, $Re = 100$): (a) Overview of the modal interaction network for fluid and structure oscillators and their inter-layer coupling. (b) Magnitude (top) and phase (bottom) of the complex supra-adjacency matrix for modal interaction. Note the inter-layer edges between structure nodes I and II and the fluid nodes (corresponding to the top of (b)) are omitted for clarity in (a). The magnitude of the edge weights are normalized with respect to the maximum edge weight for visualization in (b).}
    \label{fig:modal}
\end{figure}

\section{Results}\label{results}

\subsection{Vortical interaction network}

For the vortical interaction network described in section \ref{fsvortnet} and illustrated in Figure \ref{fig:sparsification}, we elaborate on the results in this section. 
We first look at network metrics that highlight the role of the nodes in the network in section \ref{nm}.
We then develop a data-driven model using nonlinear regression capable of predicting the community-reduced FSI vortical network structure over the limit cycle in section \ref{dbp}.
Finally, we present results from the physics-based prediction of community centroids in \ref{pbp}.

\subsubsection{Network metrics}
\label{nm}

To analyze the interactions between the fluid and structural components in the FSI system and how they change with time, we analyze the supra-adjacency network structure via network metrics. 
As we are interested in the overall inter-layer influence of the fluid on the structure and vice-versa, we construct the inter-layer supra-adjacency $\mathcal{A}_\alpha^\text{inter}$ as
\begin{equation}
    \begin{aligned}
       \mathcal{A}_\alpha^\text{inter} = 
       & \begin{bmatrix}
       \mathbf{0} & \mathcal{W}_{s\leftarrow f} \\
       \mathcal{W}_{f\leftarrow s} & \mathbf{0} 
        \end{bmatrix},
    \end{aligned}
    \label{eqn:supra_steps}
\end{equation}
where the entries on block diagonals corresponding to the structural layer and fluid layer are zero.
For the structural component, we define total $\text{out-degree} = \sum_{i}^{N_c} \sum_{j}^{n_c} \mathcal{W}_{f_i\leftarrow s_j}$ and total $\text{in-degree} = \sum_{i}^{n_c} \sum_{j}^{N_c} \mathcal{W}_{s_i\leftarrow f_j}$. 
This out-degree is the total influence of the structure on the fluid at a particular time and the in-degree is the total influence of the fluid on the structure. 
We also examine Katz centrality of the inter-layer network defined as $\bC = (\bI - \alpha \mathcal{A}_\alpha^\text{inter})^{-1} {\bf 1}$, where $\bI$ is the identity matrix, $\alpha$ is a hyper-parameter to account for nodes with zero or low eigenvector centrality, and ${\bf 1}$ is a vector of ones.
Here, we choose $\alpha = 0.01$.
To quantify the total strength of the influential community structures, we examine the measure $K = \sum_{i} C$. 

\begin{figure}[t]
    \centering
    \includegraphics[width=1.00\textwidth]{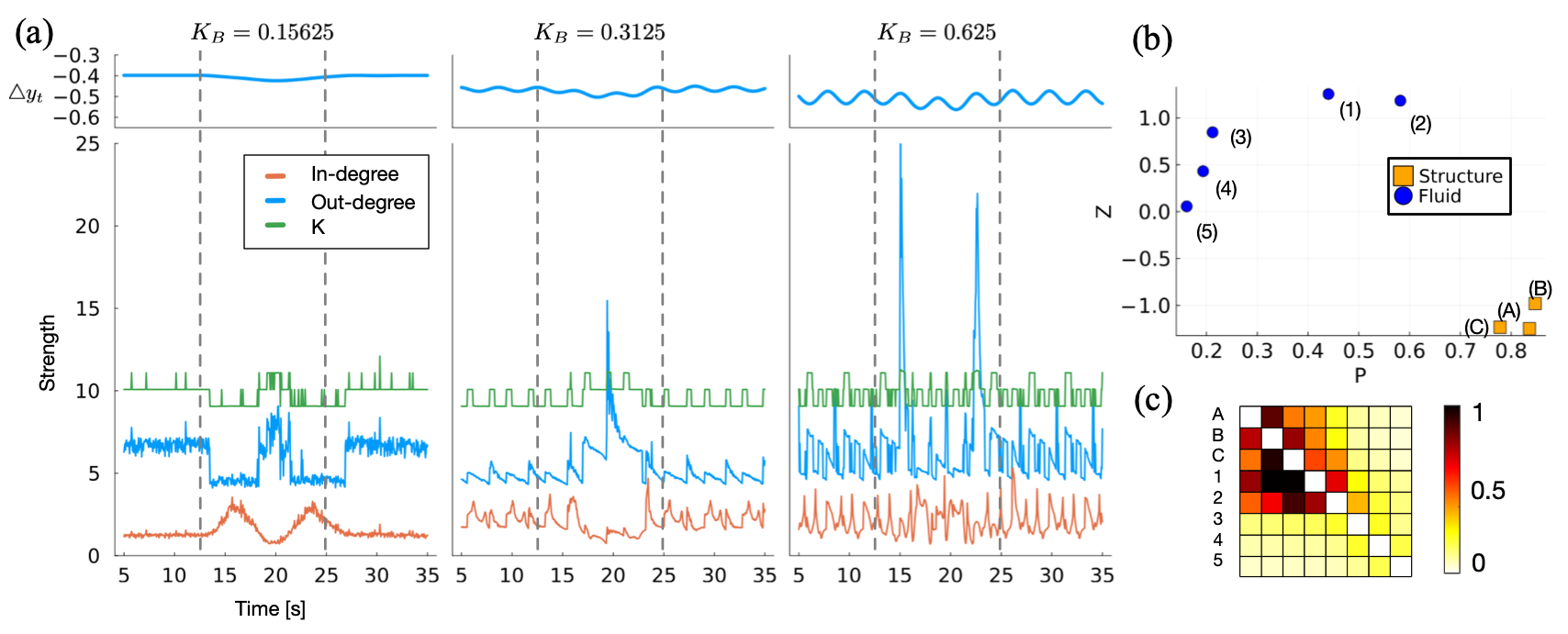}
    \caption{Network metrics for fluid-structure vortical interaction network: (a) Time evolution of centrality measures (in-degree, out-degree, and Katz measure $K$ for the structural components) for the inter-layer supra-adjacency matrix for three differnt bending stiffnesses during limit cycle and a 5-degree angle of attack gust encounter (between dashed lines). (b) P-Z map distribution of supra-adjacency matrix showing the structure nodes ($\square$) and fluid ($\circ$) and the (c) corresponding adjacency matrix. The magnitude of the edge weights is normalized with respect to the maximum edge weight for visualization in (c).}
    \label{fig:measures}
\end{figure}

We show the total in-degree, out-degree, and Katz centrality measure $K$ for each snapshot in time for the three different bending stiffness in Figure \ref{fig:measures}(a).
On the top, we show the transverse tip displacement. 
In the time between the dashed lines, a ``$1-\cos$" gust encounter is applied with a maximum pitch-down of the plate of $5^\circ$.
Limit cycle oscillations are observed at other times. 
We see that there is significant noise in degree centrality for the least compliant case of $K_B = 0.15625$.
As the structure becomes more compliant, repeating patterns in the network measures can be seen through the phase progression of the limit cycle.
The total Katz centrality measure provides the lowest noise response signal during the limit cycle.
The square wave spike and variations in $K$ indicate the detection of new communities caused by vortex shedding.

For the gust encounter, we see that the total out-degree spikes proportionally increase with compliance. 
For the low and medium compliant cases, a strong in-degree spike is observed just after the gust starts and just before it ends and a strong out-degree spike is observed during the middle of the gust encounter.
This is expected as the structure gets perturbed (influenced) during the gust encounter and as the structure deforms, it influences the rest of the flow field. 
Thus, the in- and out-degree are opposite in phase during the gust encounter for the two lesser compliant cases.
Strong out-degree spikes are seen just after the gust starts and just before it ends for the most compliant case $K_B = 0.625$.
Also, Katz centrality measure $K$ clearly detects the gust for the two lesser compliant cases; however, shows only minor changes for the most compliant case.
This indicates the changes in the vortex shedding events and formation of the new communities for the less compliant cases and not many changes in the formation of new communities for the most compliant case. 
The small amplitude of the gust and relatively slow variation gets masked by the oscillation of the structure in the most compliant case.

In addition to the network measures above, we also investigate our system using a participation score vs. z-score map (P-Z map) of the community-reduced supra-adjacency $\mathcal{A}_\alpha$. 
This provides a concise and visual depiction of the interaction characteristics of nodes within a network.
Z-score and participation coefficient are defined using the out-degree of the community-reduced supra-adjacency matrix $s_i = s_i^\text{out}$ as
\begin{equation}
    Z_i = \frac{s_{i} - \overline{s_i}}{\sigma_{s_{i}}}, ~~~~~ P_i = 1 - \left [ \left ( \frac{S^{s(f)}}{s_i} \right )^2 + \sum_{k,k \neq i} \left ( \frac{s_{k}}{s_i} \right )^2 \right ]
    \label{eqn:zscore}
\end{equation}
where $S^{s(f)}$ is the total out-degree strength of the nodes in the structure (fluid) and $\overline{s_i}$ is the mean out-degree of all centroids and $\sigma_{s_{i}}$ is the standard deviation of the out-degree strength.

The P-Z map provides an intuitive visualization of the role that each community plays in the system as seen in Figure \ref{fig:measures}(b).
The corresponding supra-adjacency matrix is shown in Figure \ref{fig:measures}(c).
Nodes with high participation scores are called connectors, while those with low-participation scores are called peripherals \cite{meena2021identifying}. 
High z-score indicates hubs that exert maximum influence within their community but have little influence over other communities. 
In fact, both peripherals and hubs do not have much inter-community influence. 
We clearly observe that all of the structure nodes have high participation scores.
Also, the centroid B which is close to the center of the plate plays the most crucial role in the interaction dynamics.
This indicates that the structural nodes are the main influencers in the FSI vortical network. 
We also see that the first two communities of the fluid have a high z-score and comparatively higher participation scores. 
These near-wake centroids have the most inter and intra-community interactions.
As communities are advected downstream we see that their influence on the structure and on the fluid diminish in a nearly linear fashion with low participation and z-score. 

\subsubsection{Data-based prediction}
\label{dbp}

\begin{figure}[h]
    \centering
    \includegraphics[width=1.00\textwidth]{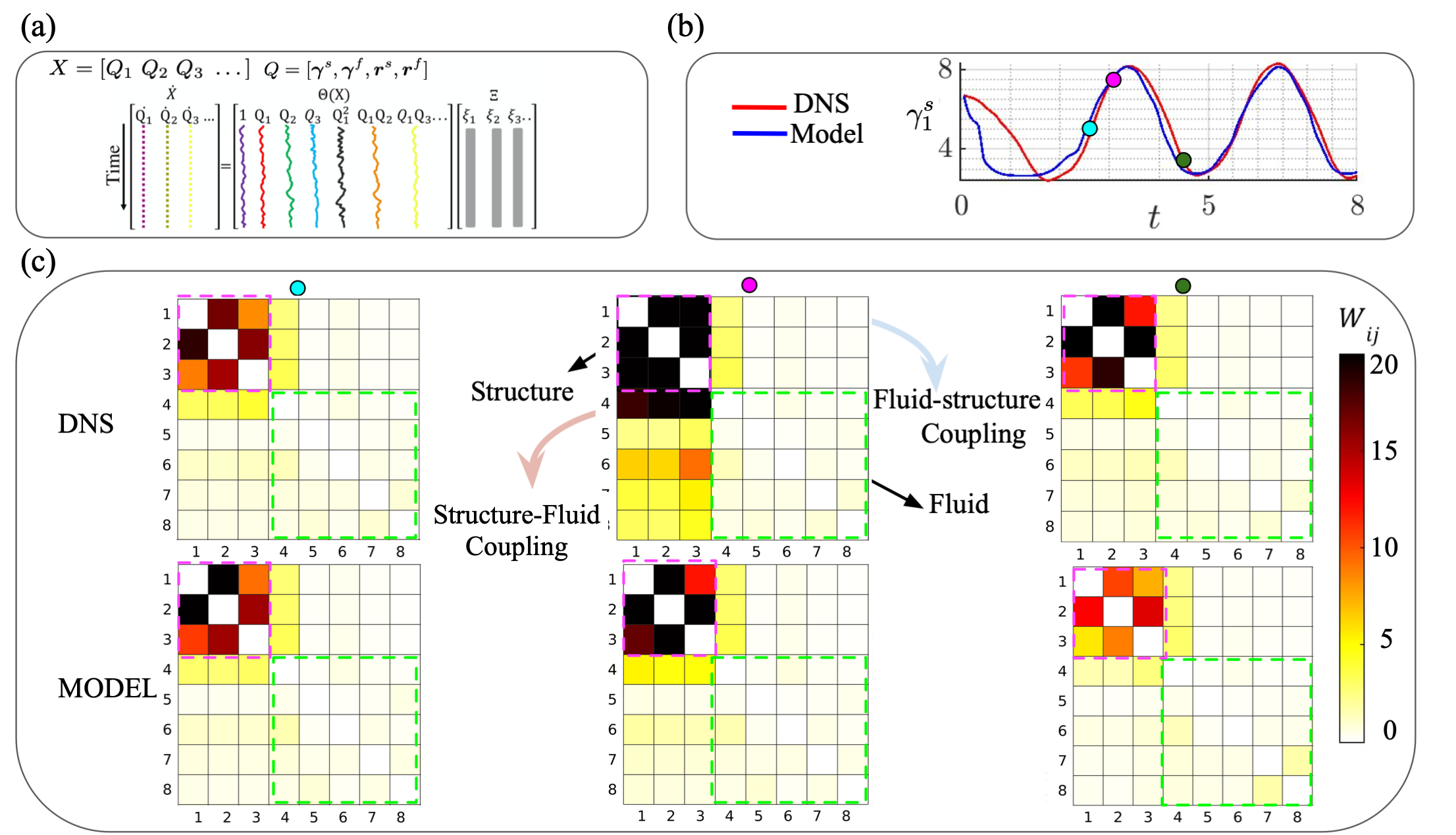}
    \caption{Data-based prediction of the fluid and structure community centroids of the vortical network: (a) Construction of the SINDy model for the evolution of circulation and position of centroids, comparison of the predicted (b) trajectories and (c) adjacency matrix of the model with that from direct numerical simulation.}
    \label{fig:potentialflow1}
\end{figure}

In this section, the time-series data of the fluid and structure community centroids $c_i$ and their associated strength $\gamma_{c_i}^{s(f)}$ and position $(x_{c_i}^{s(f)},y_{c_i}^{s(f)})$ are used to build a predictive dynamical model. 
We use sparse identification of dynamical systems (SINDy) \cite{brunton2016discovering} for generating this predictive model as shown in Figure \ref{fig:potentialflow1}(a). A sensitivity analysis of the sparsity parameter is reported in Appendix B. The values predicted by the SINDy model for the circulation of the first structure centroid compared to that from direct numerical simulation are presented in Fig \ref{fig:potentialflow1}(b). We see an acceptable agreement between the original data and the values predicted by SINDy model. The location and circulation trends for other centroids (not shown here) also match reasonably with the DNS data. 

With the SINDy model, we can now predict the evolution of the community-reduced supra-adjacency matrix as well. We show the similarity between the predicted network structure of the adjacency matrix using the model with that obtained from the direct numerical simulation at three characteristic times in Figure \ref{fig:potentialflow1}(c). This demonstrates that the relative interaction between the communities is preserved by the predictive model. The weights of edge weights are restricted to the same range to show the richness in the interactions over the limit cycle. The three structure communities exert maximum influence over the first fluid centroid corresponding to the shed positive vortical structure.

\subsubsection{Physics-based prediction}
\label{pbp}

In this section, we advect the community centroids from a single flow realization using the potential flow code developed by Darakananda et al. \cite{darakananda2016vortex}.
The plate coordinates at the time corresponding to the flow realization are provided as input to the solver.
The system is then allowed to evolve with the plate coordinates being updated at regular intervals.

\begin{figure}[h]
    \centering
    \includegraphics[width=1.00\textwidth]{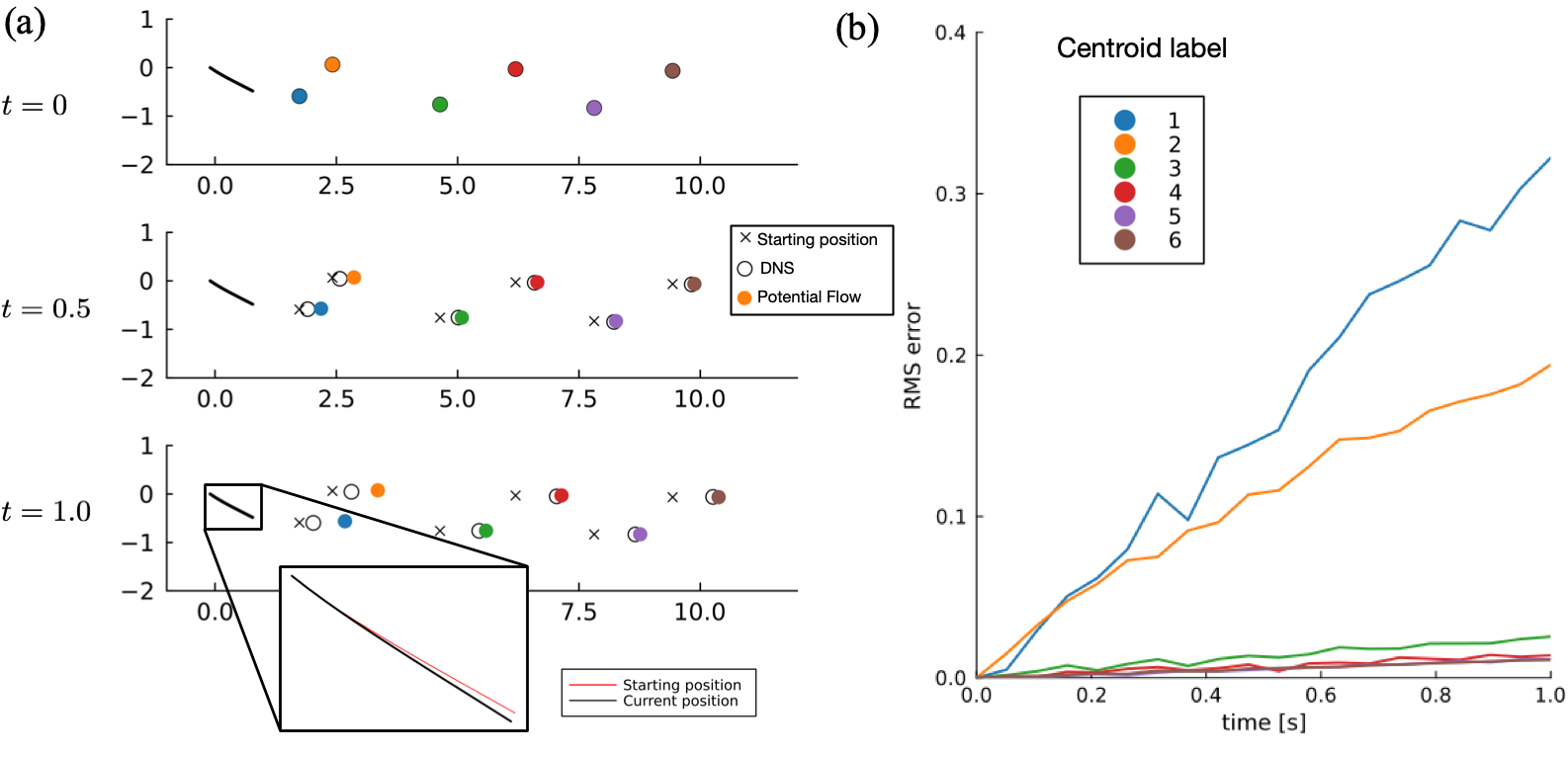}
    \caption{Physics-based prediction of the fluid community centroids of the vortical network using potential flow solver: (a) spatial position of the fluid vortical community centroids at time $t = 0$, $t = 0.5$, and $t = 1.0$ seconds.
    Inset shows the starting position, $t = 0$, and current position, $t = 1.0$, of the structure.
    (b) RMS error traces of the x-position for each of the six fluid communities compared to DNS.}
    \label{fig:potentialflow}
\end{figure}

The potential flow code is initiated at $t=0$ with the six vortices at the location of each community centroid with strengths corresponding to the total vorticity within each community.
The flow was then allowed to evolve for one second of simulation time.
The starting position of each community centroid (vortices) is denoted by an $\times$ symbol while the position of each community centroid identified from direct numerical simulation is denoted by an empty circle $\circ$.

In Figure~\ref{fig:potentialflow}(a), we show the physics-based advection of the community centroids by filled circles and compare that with those from direct numerical simulation at characteristic times. 
We see strong agreement between the physics-based advection of the seeded community centroid vortices with that of the community detection results from DNS data.
As seen in the inset of panel (a), the shape of the body is changed at regular intervals.
Figure \ref{fig:potentialflow}(b) shows the RMS error associated with the predicted x-position of each of the six vortices.
We note that the fluid communities that are closest to the structure (1) and (2) have the largest error associated with them. 
This behavior is due to the poor prediction of vortices that have not fully shed.
The leading and trailing edge suction parameter needs to be tuned when a vortex is shed \cite{narsipur2020variation}.  
The remaining four vortices in the far wake show good agreement with the DNS data.
The error in the position increases in time which can be attributed to the absence of viscosity in the potential flow solution.
Both the data-based and physics-based strategies are complementary to one another to obtain a fast prediction of FSI interactions and the dynamics of centroid communities.

\subsection{Modal interaction network}

For the modal interaction network described in section \ref{fsmodalnet} and illustrated in Figure \ref{fig:modal}, we elaborate on the results in this section. 
We first discuss the modal decomposition results in section \ref{md}.
We then discuss the results of predictions from the networked oscillator model of Eq.~(\ref{nom}) in section \ref{sec:nom}.

\subsubsection{Modal decomposition}
\label{md}

\begin{figure}[t]
    \centering
    \includegraphics[width=1.00\textwidth]{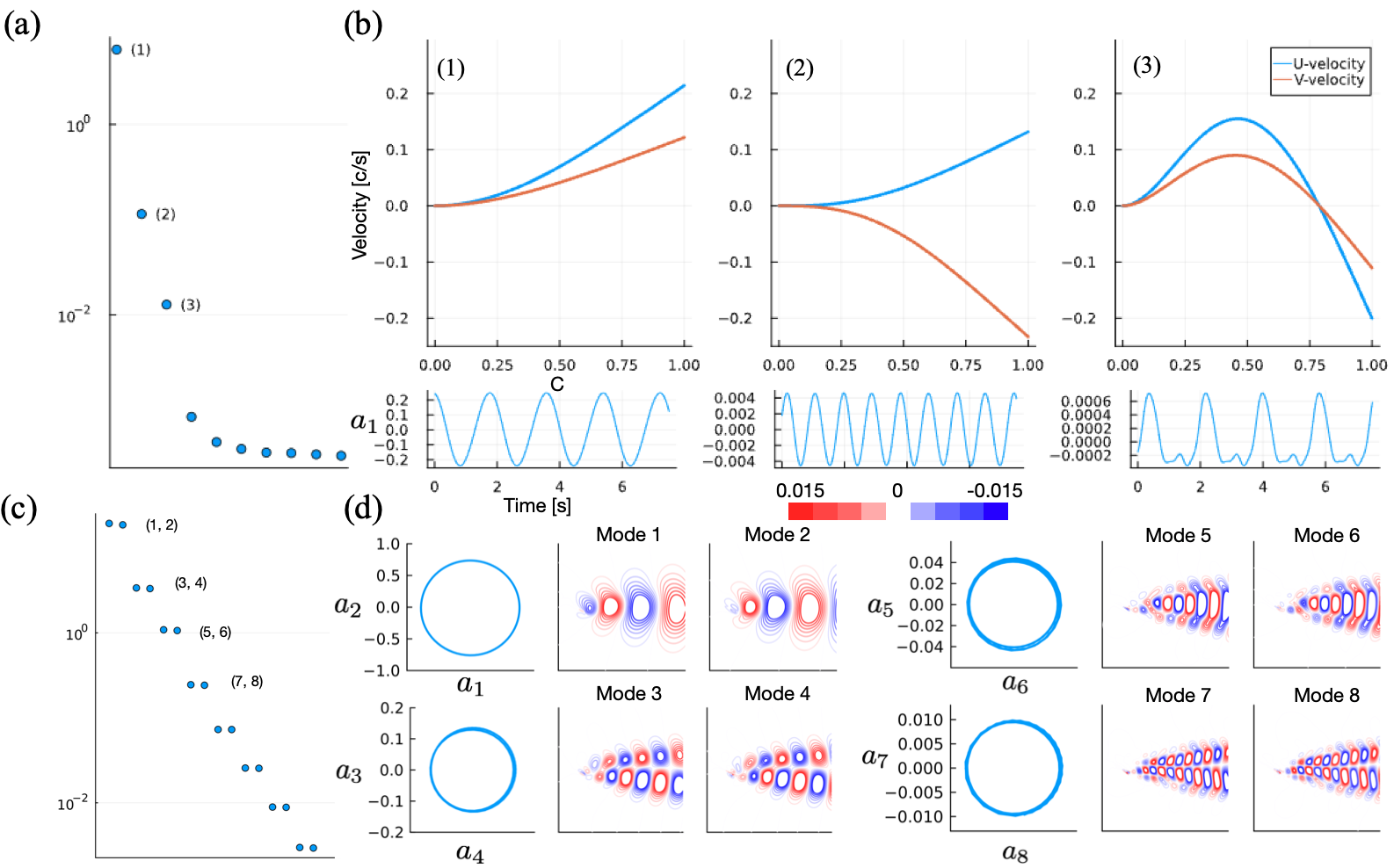}
    \caption{Modal decomposition of the fluid-structure interaction system ($M_\rho = 3$, $K_B = 0.625$, $Re = 100$). Structure layer: (a) Singular values for the first ten PCA modes of the time-series of the velocity of the structure, (b) mode shapes and temporal coefficient traces for the three structural modes selected as nodes of the modal interaction network. Fluid layer: (c) Singular values for the first eight POD mode-pairs (16 modes), (d) vorticity of the spatial modes and temporal coefficient phase portraits for the each mode-pair for the four leading mode-pairs selected as nodes of the modal interaction network.}
    \label{fig:modes}
\end{figure}

The results of the principal component analysis of the time series of velocity of the plate is shown in Figure~\ref{fig:modes}(a) and (b). 
For the structure, the singular values drop off rapidly after the third mode as seen in Figure~\ref{fig:modes}(a).
This provides a clear threshold for modal truncation. 
The mode shapes for both the x- and y-velocity component are similar to that of the bending modes of a cantilever beam, as seen in the top panel of Figure~\ref{fig:modes}(b).
We see typical sinusoidal traces of the temporal coefficients for the first two oscillators in the bottom panel of Figure~\ref{fig:modes}(b).

The singular values from the POD decomposition of the unsteady fluid velocity field snapshots are shown in Figure~\ref{fig:modes}(c). 
We choose eight fluid modes (4 mode pairs) to capture $99.9\%$ of the kinetic energy of the flow.
Phase portraits of the temporal coefficient of the POD mode pairs along with the spatial modes are shown in Figure~\ref{fig:modes}(d).
Each of the four mode-pair phase portraits shows a typical circular shape for unsteady laminar flows. 
The modal structures get smaller with increasing mode numbers and the corresponding amplitude of the temporal coefficient decreases. 

\subsubsection{Networked-oscillator model}
\label{sec:nom}

As discussed in section \ref{fsmodalnet}, we introduce different ranges of amplitude and phase impulse perturbations to the first two structural modes and collect data from direct numerical simulation. 
We perform simple regression on the data to extract the adjacency matrix $A_{mn}$ in Eq.~(\ref{eqn:oscillator_network}). 
We then evolve Eq~(\ref{nom}) to predict the amplitude and phase perturbation trajectories. 

The predicted amplitude trajectories compared to those extracted from direct numerical simulation for the three structural and four fluid oscillators (after the immediate transients in direct numerical simulation die out) are shown in Figure \ref{fig:modal}(a) and (b), respectively. 
The first two structure oscillators show excellent agreement with the simulation data.
While the third structure oscillator follows the trace of the true system, it has a high-frequency oscillation throughout the 50 seconds of simulation time.
This high-frequency vibration is possibly due to the low amplitude associated with the third structural oscillator.
The fluid oscillators also show comparable agreement, however, the results deviate slightly for fluid oscillator IV.
Similar agreement is observed in the phase of the perturbations (not shown). 

\begin{figure}[t]
    \centering
    \includegraphics[width=1.00\textwidth]{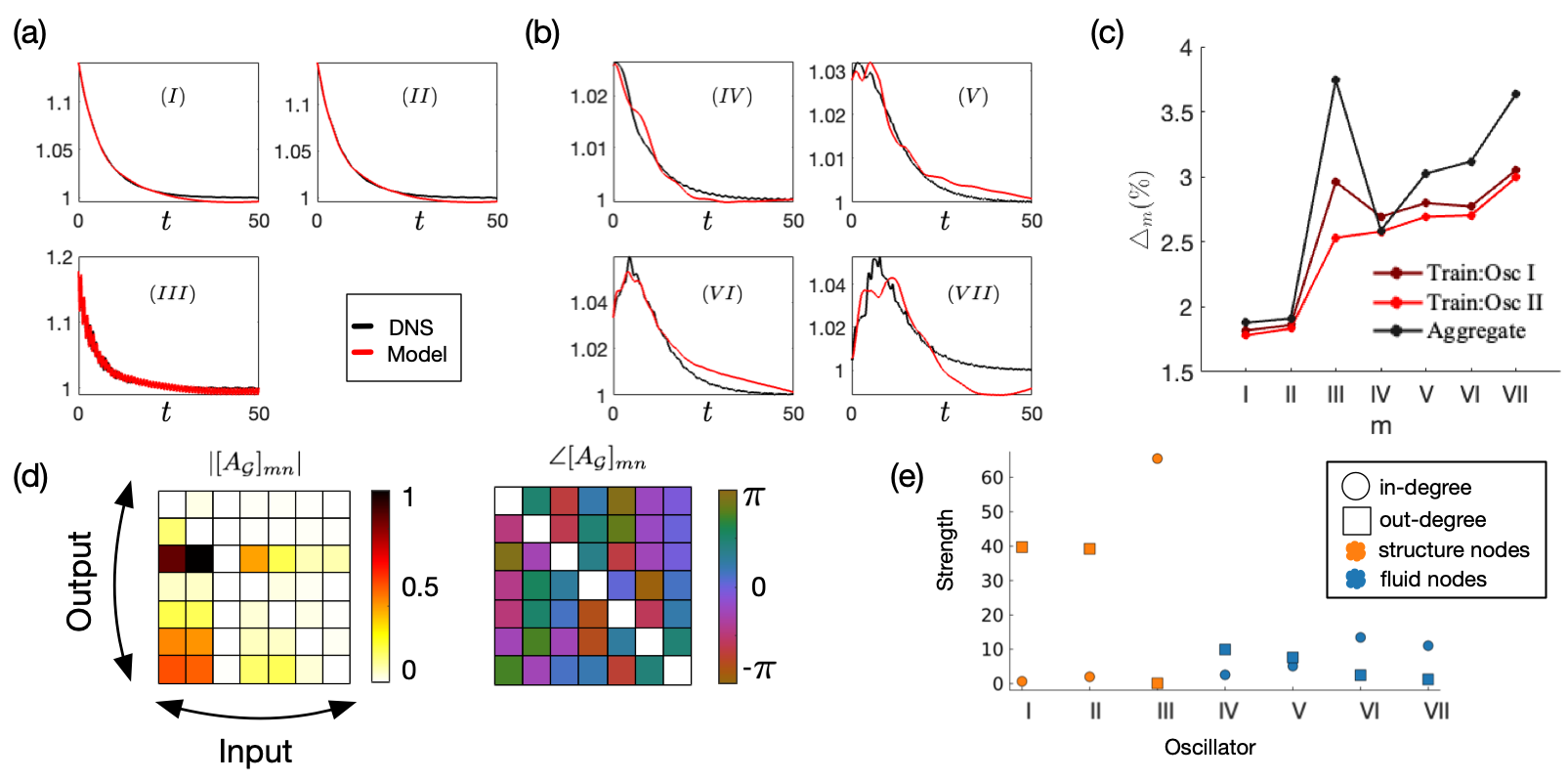}
    \caption{Network-oscillator model for fluid-structure interaction system ($M_\rho = 3$, $K_B = 0.625$, $Re = 100$): Trajectories of the three (a) structure and four fluid (b) oscillators for the predictive model (red) and ground truth (black) for $50$ seconds,
    (c) performance of the single-oscillator-based models and the aggregate model (black).
    (d) modal interaction model magnitude and phase adjacency matrices after training.
    (e) In- and out-degree for the network nodes.}
    \label{fig:modal}
\end{figure}

We extract different network models; only considering data from perturbations on structure oscillator I, only considering data from perturbations on structure oscillator II, and training from both perturbations. $20\%$ of the data from all perturbation cases are reserved for testing. 
For each of the training epochs, perturbations on structure oscillator II shows the best agreement with the test data while oscillator one shows only a slight increase in error.
The aggregate model shows the poorest performance, especially in structure oscillator III. 
All models show similar errors for the two dominant structure modes and the dominant fluid mode pair. 

The amplitude and phase relationship between the modal oscillators of the FSI system is shown in Figure \ref{fig:modal}(c).
The network structure captures the energy transfers between the modes of the structure and fluid on the introduction of impulse perturbations. 
The associated network centrality measures are shown in Figure \ref{fig:modal}(d).
The first two structure oscillators have the highest out-degree  while the third structure oscillator has the highest in-degree.
The out-degree for the fluid oscillators decrease with oscillator number while the in-degree increases.
These results for the fluid oscillators are in agreement with that of Nair et al. \cite{nair2018networked}.

\pagebreak 
\section{Conclusion}\label{conclusion}

In summary, we develop two reduced-order models of fluid-structure interaction, leveraging a multi-layer network framework. 
The two approaches use distinctive vortical and modal features of the overall FSI system.
In the vortical approach, the nodes of the fluid layer are represented by the vorticity associated with each grid cell in the Eulerian computational domain. On the other hand, the nodes of the structural layer are formed by bound vortexlets.
The edge weights in this approach are defined using induced velocity. 
Community detection was used to construct a reduced representation of the vortical network. 
In the second approach, coherent modes from the fluid and structure form the nodes of the network. 
A data-driven approach is then utilized to construct the modal interaction network model after introducing impulse perturbation to the structural modes and tracking the amplitude and phase of the modal perturbations.

Two-dimensional flow over a compliant flat plate at an angle of attack $\alpha = 35^\circ$ was investigated using the network-based approach.
Data from direct numerical simulations of three different plate stiffnesses during the limit cycle and gust encounters were converted to a community-reduced vortical network.
The network metrics were able to capture the dynamics of the limit cycle and the influence of gust encounters. 
A P-Z map was constructed to illustrate the unique role of each node of the vortical network in the overall FSI system.
Prediction of vortex dynamics and the network interactions were performed using two different strategies: a pure data-based strategy using SINDy and a physics-based strategy using a potential flow solver which was initialized using the data of community centroids.
Both methods show acceptable agreement between the prediction and ground truth data.

Subsequently, we illustrate the process of extracting the modal-interaction network the most compliant structure, $K_B = 0.625$.
Using principal component analysis of the velocities of the structure and proper orthogonal decomposition of the fluid velocity fields, modal representations for the network were obtained.
Oscillators are formed from the fluid conjugate mode-pairs and a Hilbert transform of the structural temporal coefficients.
The two dominant structural modes are perturbed to track the energy transfer in the FSI system.
We then train our network model with 80\% of the perturbation data using simple regression.
Oscillator amplitude trajectories are predicted from the model and showed close agreement with the retained testing data.

We see the possibility for this formulation to be extended into several areas. 
Firstly, it can be applied to investigate the interactions between multiple bodies in an unsteady fluid flow such as that occurring between the main wing and empennage of an airplane.
Secondly, it has the potential to facilitate the development of a computationally efficient predictive model that is well-suited for online control applications, specifically for addressing gust alleviation requirements.
Lastly, this approach offers a generalizable framework for characterizing and modeling multiphysics systems, enabling its application to a wide range of engineering disciplines.

\section*{Acknowledgements} 

AGN acknowledges the support from the Department of Energy Early Career Research Award (Award no: DE-SC0022945, PM: Dr. William Spotz) and the National Science Foundation AI Institute in Dynamic systems  (Award no: 2112085, PM: Dr. Shahab Shojaei-Zadeh).

\section*{Appendix A: Grid convergence study}\label{secA1}

\begin{figure}[h]
    \centering
    \includegraphics[width=1.00\textwidth]{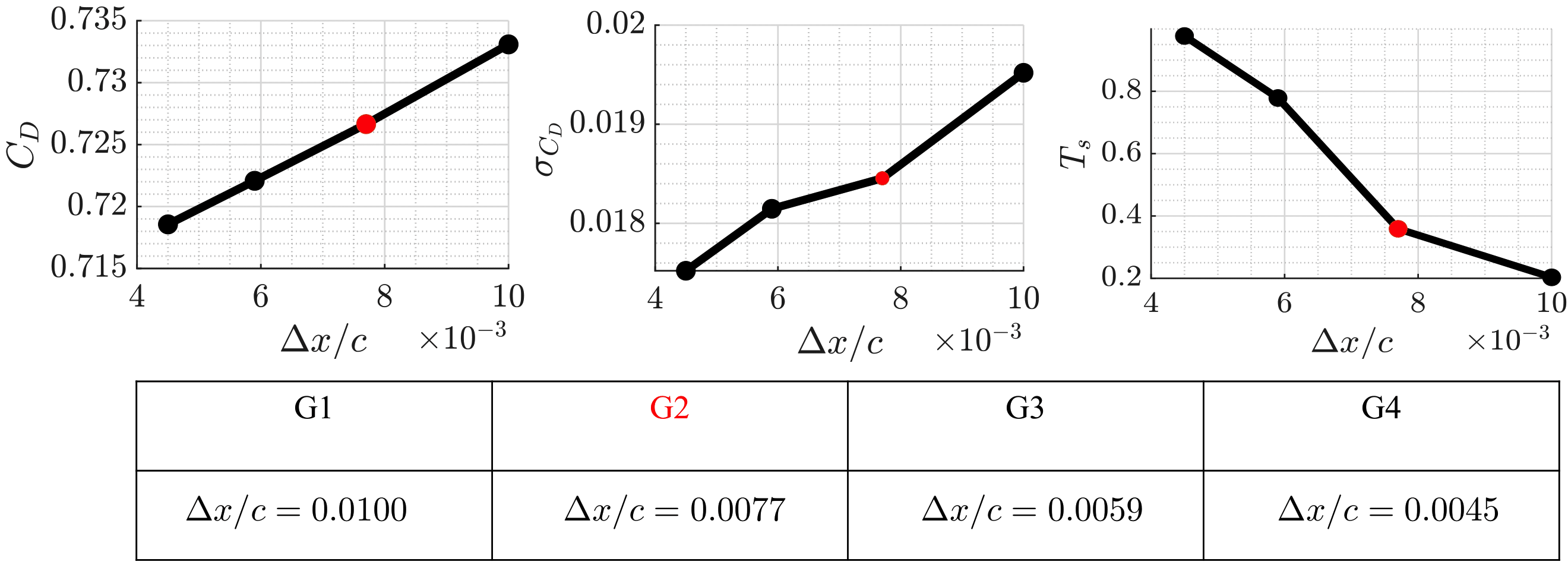}
    \caption{Grid convergence study of direct numerical simulation of a flow past a flexible flat plate at Reynolds number $Re = 100$, angle of attack $\alpha = 35^\circ$, mass ratio $M_\rho = 3$, and stiffness $K_B = 0.3125$. The variation of mean drag coefficient (left), standard deviation of drag coefficient (middle) and time required for one time step (right) with different grid spacing is shown. The grid resolution marked in red used in this work achieves a good compromise in simulation time and accuracy.}
    \label{fig:1}
\end{figure}

\begin{figure}[h]
    \centering
    \includegraphics[width=0.35\textwidth]{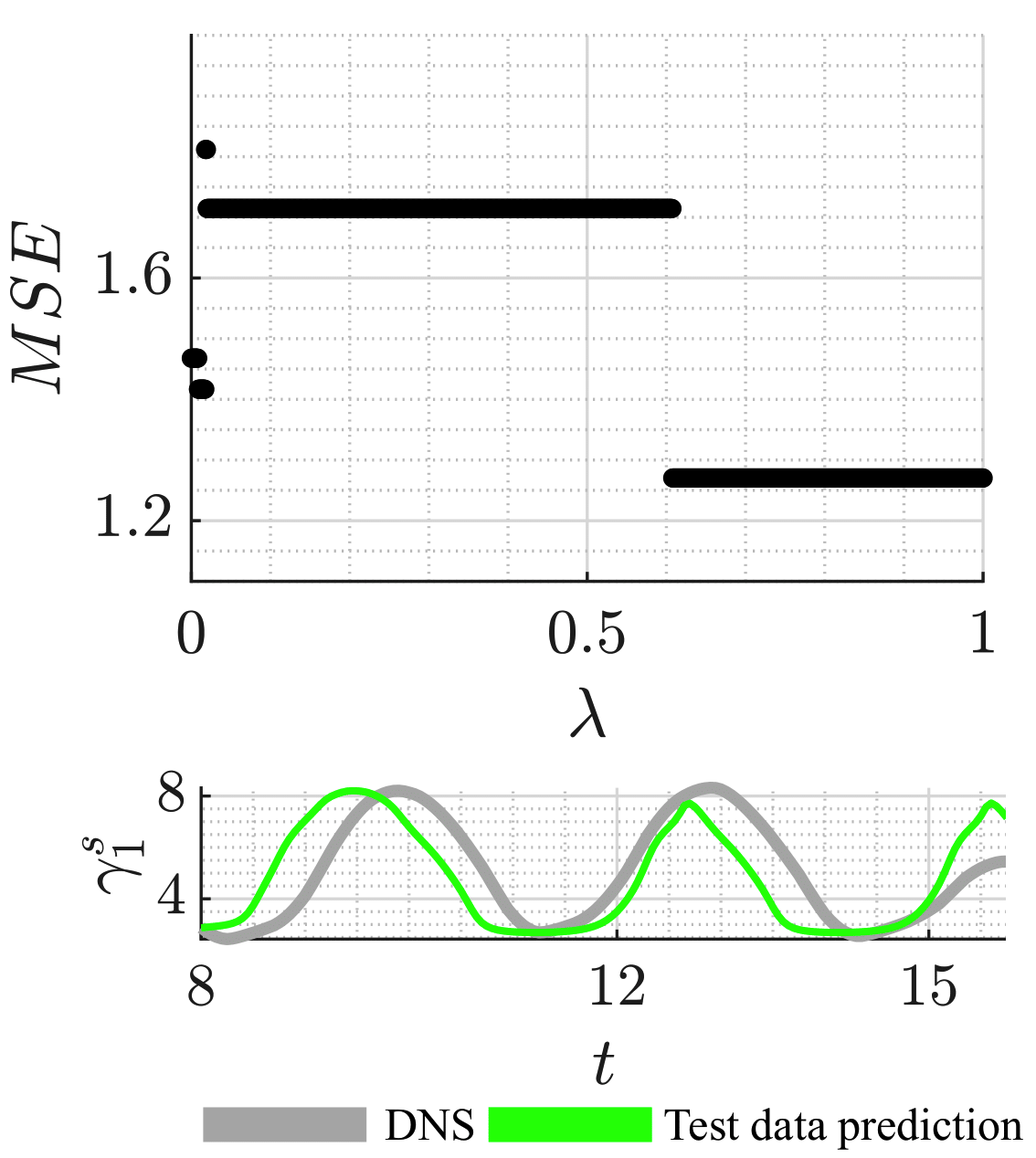}
    \caption{(top) The mean square error in prediction of the test data with different models of sparsity in SINDy. (bottom) The prediction of the vortexlet strength trajectory compared to DNS for $\lambda > 0.6$.}
    \label{fig:2}
\end{figure}

The immersed boundary projection methodology used in our study is based on the work by Taira and Colonius \cite{taira2007immersed} and Goza and Colonius \cite{goza2017strongly}. It has been applied in numerous fluid-structure interaction investigations \cite{hickner2022data,nair2022fluid,goza2018modal}. We conducted a grid convergence analysis focusing on a Reynolds number of $Re = 100$, an angle of attack $\alpha = 35^\circ$, a mass ratio $M_\rho = 3$, and a bending stiffness $K_B = 0.3125$. This was accomplished by implementing four different grid setups.

Our solver employs a multi-domain approach to expedite the computations \cite{colonius2008fast}. The grid spacing we discuss here pertains to the innermost domain featuring the finest grid. We illustrate the mean drag coefficient ($C_D$), its standard deviation, and the duration of a single iteration for each grid in Fig. \ref{fig:1}. Additionally, Fig. \ref{fig:1} shows the grid spacing for the various grids.

In our research, we use the grid denoted as $G2$ for simulations, with a corresponding $\Delta x/c=0.0077$, notably highlighted in red. The subsequent level of refinement, grid $G3$, results in a mere $0.6\%$ alteration in the mean $C_D$. The standard deviation for $C_D$ registers at $0.01845$ for grid $G_2$ and slightly less at $0.01815$ for grid $G_3$. Notably, the computational time for a single iteration on grid $G_3$ nearly doubles that of grid $G_2$. Consequently, grid $G_2$ delivers acceptable precision and shorter computational time, making it our choice for all simulations in this case, with a grid spacing of $\Delta x/c=0.0077$.

\section*{Appendix B: Sensitivity of sparsity parameter}\label{secA2}%

Sparse identification of nonlinear dynamics (SINDy) \cite{brunton2016discovering} employs an L$_1$ regularization which allows to construct a sparse model for the resulting dynamics. We present a sensitivity analysis of sparsity promoting factor $\lambda$ by first dividing the data set into training and test data. The mean square error in the prediction of test data set for different $\lambda$ is presented in Fig. \ref{fig:2}. It is observed that $\lambda > 0.6$ produces minimum error for test data prediction. We employed a value $\lambda=0.6$ for the data-based prediction in the original manuscript.

\bibliographystyle{unsrt}
 \bibliography{bibliography}

\end{document}